\documentclass[10pt,letterpaper]{article}
\usepackage[top=0.85in,left=2.75in,footskip=0.75in]{geometry}

\usepackage{amsmath,amssymb}
\usepackage{hyperref}% add hypertext capabilities
\hypersetup{colorlinks=true}
\usepackage{changepage}

\usepackage[utf8]{inputenc}

\usepackage{textcomp,marvosym}

\usepackage{nameref,hyperref}

\usepackage[right]{lineno}

\usepackage{microtype}
\DisableLigatures[f]{encoding = *, family = * }

\usepackage[table]{xcolor}

\usepackage{array}

\usepackage[citestyle=numeric, backend=biber,sorting=none]{biblatex}
\newcolumntype{+}{!{\vrule width 2pt}}

\newlength\savedwidth

\raggedright
\usepackage[aboveskip=1pt,labelfont=bf,labelsep=period,justification=raggedright,singlelinecheck=off]{caption}

\makeatletter
\renewcommand{\@biblabel}[1]{\quad#1.}
\makeatother

\usepackage{lastpage,fancyhdr,graphicx}
\usepackage{epstopdf}
\fancyheadoffset[L]{2.25in}
\fancyfootoffset[L]{2.25in}
\newcommand{\totalN}{160,933}

\begin{document}
\vspace*{0.2in}

\begin{flushleft}
{\Large
\textbf\newline{Predicting time to graduation at a large
enrollment American university} % Please use "sentence case" for title and headings (capitalize only the first word in a title (or heading), the first word in a subtitle (or subheading), and any proper nouns).
}
\newline
\\
John M. Aiken\textsuperscript{1,3}
Riccardo De Bin\textsuperscript{2}
Morten Hjorth-Jensen\textsuperscript{1, 3, 4}
Marcos D. Caballero\textsuperscript{1, 3, 5}

% \\
\bigskip

\textbf{1} University of Oslo Department of Physics, Centre for Computing in Science Education, Blindern, Oslo, Norway
\\
\textbf{2} University of Oslo Department of Mathematics and Statistics, Blindern, Oslo, Norway
\\
\textbf{3} Michigan State University Department of Physics and Astronomy, East Lansing, Michigan, USA
\\
\textbf{4} Michigan State University, National Superconducting Cyclotron Laboratory and Facility for Rare Ion Beams, East Lansing, Michigan, USA
\\
\textbf{5} CREATE for STEM Institute, Michigan State University, East Lansing, Michigan 48824
\\
\bigskip

% Insert additional author notes using the symbols described below. Insert symbol callouts after author names as necessary.
% 
% Remove or comment out the author notes below if they aren't used.
%
% Primary Equal Contribution Note
% \Yinyang These authors contributed equally to this work.

% Additional Equal Contribution Note
% Also use this double-dagger symbol for special authorship notes, such as senior authorship.
% \ddag These authors also contributed equally to this work.

% \textcurrency Current Address: Dept/Program/Center, Institution Name, City, State, Country % change symbol to "\textcurrency a" if more than one current address note
% \textcurrency b Insert second current address 
% \textcurrency c Insert third current address

% \dag Deceased

% Group/Consortium Author Note
% \textpilcrow Membership list can be found in the Acknowledgments section.

% Use the asterisk to denote corresponding authorship and provide email address in note below.
* j.m.aiken@fys.uio.no

\end{flushleft}
\section*{Abstract}
The time it takes a student to graduate with a university degree is mitigated by a variety of factors such as their background, the academic performance at university, and their integration into the social communities of the university they attend. Different universities have different populations, student services, instruction styles, and degree programs, however, they all collect institutional data. This study presents data for \totalN $\;$ students attending a large American research university. The data includes performance, enrollment, demographics, and preparation features. Discrete time hazard models for the time-to-graduation are presented in the context of Tinto's Theory of Drop Out. Additionally, a novel machine learning method: gradient boosted trees, is applied and compared to the typical maximum likelihood method. We demonstrate that enrollment factors (such as changing a major) lead to greater increases in  model predictive performance of when a student graduates than performance factors (such as grades) or preparation (such as high school GPA).

% Please keep the Author Summary between 150 and 200 words
% Use first person. PLOS ONE authors please skip this step. 
% Author Summary not valid for PLOS ONE submissions.   
% \section*{Author summary}
% Lorem ipsum dolor sit amet, consectetur adipiscing elit. Curabitur eget porta erat. Morbi consectetur est vel gravida pretium. Suspendisse ut dui eu ante cursus gravida non sed sem. Nullam sapien tellus, commodo id velit id, eleifend volutpat quam. Phasellus mauris velit, dapibus finibus elementum vel, pulvinar non tellus. Nunc pellentesque pretium diam, quis maximus dolor faucibus id. Nunc convallis sodales ante, ut ullamcorper est egestas vitae. Nam sit amet enim ultrices, ultrices elit pulvinar, volutpat risus.

% \linenumbers
\section{Introduction}

University students must meet a number of objectives to obtain degrees and in many cases this can prolong their time at the university \parencite{yue2017rethinking} or they drop out altogether \parencite{braxton2000influence}. During their studies, American students may take on substantial financial obligations when choosing to pursue degrees and extending beyond the “four-year degree” can greatly increase the cost of obtaining that degree \parencite{abel2014recent}. Thus understanding the paths that students take towards degree completion can help faculty and administrators better serve student populations to meet their educational goals. Tinto's Theory of Drop Out \cite{tinto1975dropout} has seen large acceptance in its ability to describe the factors that influence a student's path towards degree completion \parencite{pascarella1983predicting, nora1990testing, cabrera1993college, desjardins1999event, ishitani2003longitudinal}. Tinto theorized that a student's college drop out decision is mediated by two conglomerate features: 1) educational goal commitment, and 2) institutional commitment. He further theorized that these commitments are dynamic. Students begin their university studies with initial commitments that are then mediated by how students participate in the academic and social systems of the university. Tinto suggested that using institutional data such as that collected by university registrars, could quantify these relationships and provide predictive models for determining student's at-risk to drop out or alternatively, graduate.

This paper examines student ultimate success at university. We define this success as obtaining a bachelor's degree. It uses 20 years of institutional data for \totalN $\;$students attending a large enrollment American research university. This paper examines two specific research questions:

\begin{enumerate}
    \item In the context of Tinto's Theory of Drop Out \cite{tinto1975dropout}, what factors (such as grades, participation in a major, student background, etc.) contribute to the time it takes to obtain a degree at a large enrollment research university in the United States? How does the contribution of these factors change for longer durations until graduation?
    \item How can recent innovations in statistics and machine learning, such as gradient boosting and xgboost, improve educational model performance?
\end{enumerate}

We focus on a comparison of student participation in an academic system, their involvement in the social system, and their initial conditions due to high school training. We demonstrate that enrollment factors (e.g., changing a major) and academic performance are more important to predicting when a student graduates than pre-college experiences (e.g., high school GPA). Additionally, we compare traditional statistical modeling (maximum likelihood estimation) to new techniques from machine learning (gradient boosting \parencite{chen2016xgboost}) and demonstrate that the machine learning methods are more effective at estimating the function predicting time-to-graduation.

It is important to use an analysis technique that respects the dynamics assumptions of Tinto's Theory of Drop Out. In this paper we use a Discrete Time Hazard Model framework \parencite{yamaguchi1991event}. Discrete time hazard model is useful when the classic Cox regression model assumption that events happen on a continuous interval no longer hold valid \parencite{yamaguchi1991event}. This is true when data has highly discretized time intervals such as semesters enrolled. Traditionally this modeling is done with logistic regression via maximum likelihood estimation \parencite{yamaguchi1991event}. In this paper we compare two models that fit the logistic model including logistic regression (maximum likelihood estimation) and a gradient boosted tree model (xgboost, \cite{chen2016xgboost}).  
% Additionally, a discrete time hazard model framework follows the supposition by Tinto that a student's intent to graduate is dynamic and changes over time based on at-college experiences.

\section{Background}

% This paper grounds its predictive models in Tinto's theory of dropout \parencite{tinto1975dropout}. It uses a discrete time hazard model (discrete time hazard model) \parencite{yamaguchi1991event} that respects the dynamic assumptions of Tinto's theory. And it compares the statistical methods that are traditionally used to predict duration event data (e.g., \cite{desjardins1999event, ishitani2003longitudinal, chen2012institutional, yue2017rethinking}) in education research with new machine learning algorithms \parencite{chen2016xgboost}. 

In this section Tinto's theory\parencite{tinto1975dropout} and the use of that theory will be described, the use of discrete time hazard model \parencite{yamaguchi1991event} to predict when students graduate and/or drop out  will be described, and the value of a new machine learning method known as gradient boosting will be described \parencite{chen2016xgboost}. The functional descriptions of the discrete time hazard model and gradient boosting are described in Section \ref{sec:methods}.

% \textcolor{red}{strong summary of tintos theory of drop out along with other references using it}\\
\subsection{Tinto's Theory of Drop Out}

Tinto's Theory of Drop Out describes a students intent to drop out based on the interplay of two quantities: 1) a student's commitment to education, and 2) a student's commitment to a specific institution. A student's commitment to education is mediated by the initial state of a student entering the university and the dynamics that occur while the student attends university. These dynamics are dictated by a student's participation and acceptance into the social and academic communities that are at a university. A student's commitment to an institution is tempered by many factors such as the educational goals available at an institution (e.g., a technical university degree offerings versus a liberal arts university), family commitment to a university, and social acceptance at the university. While Tinto's theory explicitly attempts to describe why students drop out from college, it is not uncommon that this theory is used to study student graduation from college given that graduation is an alternative outcome to dropping out \parencite{ishitani2003longitudinal, scott2005pitfalls, yue2017rethinking}. In this paper we focus on a student's commitment to education as the framing for features that predict when a student will graduate if they do so.

% \cite{tinto1975dropout} Theory of Drop Out's identifies several contributing factors to a student's educational commitment. These factors include initial conditions such as a students demographics and preparation for university. They also include dynamic factors such as a student's integration into the university community of their chosen degree program and success in their chosen degree.

It is common to characterize Tinto's theory of dropout as a discrete time hazard problem \parencite{desjardins1999event, ishitani2003longitudinal, scott2005pitfalls, chen2013stem, yue2017rethinking}. Tinto's theory posits that a student's educational commitment changes over time as they work toward's a degree and that effects, such as high school GPA, that may be strong in the beginning of a degree program are weak toward's the end of a degree program. Discrete time hazard modeling provides a systematic method for examining the various effects on the probability to graduate over time \parencite{yamaguchi1991event}. To that end, discrete time hazard modeling has been used to examine the dynamic impact of various effects that Tinto predicts effect a student's educational commitment. College GPA has been demonstrated to have a profound but diminishing time varying effect on graduation \parencite{desjardins1999event, chen2012institutional, yue2017rethinking}. Financial aid and money spent on student services has also demonstrated a time varying (diminishing) effect on the probability to graduate \parencite{chen2012institutional, yue2017rethinking}. First generation student's are considered high risk for drop out but this effect diminishes over time indicating that first generation students need specific resources other students may not \parencite{ishitani2003longitudinal}. Non-traditional enrollment factors such as delaying enrollment, working while enrolled, and stopping education for some period of time can all have a negative time varying effect on graduation \parencite{scott2005pitfalls}.
% Student's who receive financial aid  demonstrated that first generation students are more likely to depart college in comparison to students with college educated parents but that this effect diminishes as the student stays enrolled. \cite{desjardins1999event} found that while GPA is a primary effect in continued enrollment at the beginning of the college career, this effect wanes as students continue enrollment towards graduation. Using national data for 400 institutions, \cite{chen2012institutional} confirmed Tinto's supposition that preparation and college experience both impact a students commitment to stay in education and also suggests that the money that is spent on student services is a predictor for student retention. 
In each case these studies showed a statistically significant time varying impact that supports the dynamic claims of Tinto's theory.

A student's preparation has long been known to impact a student's ability to graduate. Preparation is assessed in many ways and can be represented as the experiences a student has in school and also their present innate ability to perform a specific function. Math preparation correlates with both performance in university and graduating \parencite{trusty2003high, gaertner2014preparing}. The same is true for physics and english preparation \parencite{bamberg1978composition, sadler2001success, hazari2007gender}. High school GPA and SAT scores typically account for some but not all of student success at university (GPA, etc.) \parencite{zwick2005predicting, aiken2019modeling, zabriskie2019using}. Preparation for university often can be experienced differentially as well. Women in STEM (Science, Technology, Engineering, and Mathematics) degree programs report having less access to laboratory experiences in high school, are encouraged towards science by their father's differently, and overall have a different preparation than their male counterparts \parencite{seymour2000talking, hazari2007gender}. 

A student's demographics can include the student's gender, race, the financial support they can expect from their family, and if a student is the first in their family to attend college. Race has long been shown to be a factor in whether a student graduates or not. Black students are less likely to graduate than fellow white students when adjusted for socio-economic status and academic ability and they are more likely to have unwelcoming experiences in STEM programs while at university \parencite{alexander1982social, seymour2000talking, yue2017rethinking}. Female students are also likely to have unwelcoming experiences that cause them to switch from STEM programs \parencite{seymour2000talking}. However they have been show to be equally likely to graduate or more likely to graduation in comparison to their male counterparts \parencite{alexander1982social, seymour2000talking}. The financial support both in terms of loans and scholarships and the socio-economic status of a student's family have long been known to be a factor in university graduation with students who have more financial support typically being more likely to graduate \parencite{alexander1982social, sewell1967socioeconomic, smith2001factors, rowan2007predictors, yue2017rethinking}. First generation students are less likely to participate in extra-curricular activities at university \parencite{choy2001students, pascarella2004first} and are more likely to dropout even when adjusting for race, family income, gender, and preparation \parencite{ishitani2003longitudinal}. Ultimately, American students from different backgrounds often see different success rates at university due to the different experience they have at the university due to their race, gender, or socio-economic status. 

\subsection{Gradient boosting}

% In this paper we use a novel method for solving logistic problems: gradient boosting.

In this paper we use a novel method known as "gradient boosting". Gradient boosting can solve a large number of statistical modeling problems including logistic problems as found in this paper. Logistic problems are a group of statistical problems that assume that a sigmoidal function made from parameters $\theta$ and data $\boldsymbol{x}$ (i.e., $[1+e^{-\theta^{T} \boldsymbol{x}}]^{-1}$) approximates the probability $P(Y|X)$. In plain language, the model uses input data $X$ to determine how likely the outcome, $Y$, is to be true. In the context of this paper that would mean the input data $X$ is used to predict whether a student will graduate in the following semester. In education research, the maximum likelihood method has commonly been used to find the solutions to logistic problems. Maximum likelihood solves the logistic regression problem by picking parameters $\theta$ that maximize the log likelihood function \parencite{hastie2005elements}.
% :

% \begin{equation}
%     \frac{\delta LL(\theta)}{\delta \theta_{j}} = \sum_{i=1}^{n}[y^{(i)} - \frac{1}{1+e^{-\theta^{T} \boldsymbol{x} }}] x_{j}^{(i)}
% \end{equation}

Since there is no closed form solution to this likelihood equation the solution is found iteratively via some optimization algorithm \parencite{hastie2005elements}. Maximum likelihood is the typical default solution for logistic problems in most statistical packages such as statsmodels \parencite{seabold2010statsmodels} in python and glm in R \parencite{glmr}.

% Gradient boosting is a method to produce the solution to the logistic problem as well. Gradient boosting is a method to fit models that combines two approaches: fitting residuals and boosting. Boosting is the process of additively combining many weak learners to produce an overall model \parencite{hastie2005elements}. The weak learners learn specific rules in the local data neighborhood that combine to form a model with better predictions that are not over-fit \parencite{freund1999short}.  Fitting residuals is the process of instead of fitting on raw data, a model is fit on residuals from some null model (e.g., the average) \parencite{hastie2005elements}. Thus, gradient boosting fits a model by sequentially fitting small highly biased models to residuals passed forward from each iteration starting from the null model \parencite{friedman2001greedy}. An early addition to gradient boosting was stochastic gradient boosting that uses column and row subsampling to reduce model bias \parencite{friedman2002stochastic}. The sum total effect is a model that avoids over-fitting while reducing the bias and variance in the predicted outcome variable \parencite{suen2005combining}. 

Gradient boosting produces the solution to the logistic problem in a different way. The log-likelihood is maximized (or, in machine learning terms, the negative log-likelihood is minimized) in small steps: at each step, called boosting iteration, the gradient of the log-likelihood is computed to identify the best direction for the maximization, and it is then fed to a base learner, through which the model is updated. Different choices of base learners (in this paper we will use trees) provide different solutions, guaranteeing maximum flexibility \cite{buhlmann2007boosting}. An early stop, through a tuning parameter which controls the number of iterations, prevents overfitting and provide a better bias-variance trade-off \cite{mayr2014extending}. If the base learner is weak enough in comparison to the signal-to-noise, it can be shown that, at least for a continuous response, boosted models outperform their unboosted versions in term of MSE \cite{buhlmann2003boosting}. Several modifications of the boosting algorithm have been proposed, including stochastic gradient boosting, that uses column and row subsampling to further avoid overfitting and even better deal with the bias-variance trade-off issue \cite{suen2005combining}.

Gradient boosting has been demonstrated to produce better fit models in comparison to traditional methods across a number of domains both in binary classification and in hazard modeling \parencite{lombardo2015binary, xu2017pdc, ma2017prioritizing}. Boosted logistic models have been demonstrated to be more effective at fitting data than models that use maximum likelihood estimation \parencite{de2017data}. The structure of educational data can be highly complex often containing many sub groups that have different effects \parencite{bryk1989toward}. It is likely that gradient boosting can provide better parameter estimation for education research questions as well.

In this paper we use the Extreme Gradient Boosting (xgboost) algorithm \parencite{chen2016xgboost}. Xgboost implements stochastic gradient boosting \parencite{friedman2002stochastic} with column and row-wise subsampling, regularization, decision tree base learners with a custom tree split finding algorithm, and an in-model data imputation system for missing data. The specifics of xgboost are explained in Section \ref{sec:methods}.

\section{Data Set}

% /****** This query produces the statistic below for the <=4 semesters  ******/
% DECLARE @semidx TABLE (id varchar(9), semester_idx int)
% INSERT INTO @semidx
% SELECT [student_id_fk], MAX(semester_idx) as semester_idx
%   FROM [msu_v1].[dbo].[un_majors_aggregates]
%   GROUP BY student_id_fk
% SELECT SUM(semester_idx)
% FROM @semidx
% SELECT semester_idx, COUNT(id) * 1.0/ (SELECT COUNT(DISTINCT student_id_fk) FROM un_majors_aggregates)
% FROM @semidx
% GROUP BY semester_idx

The data in this study comes from registrar information from a large enrollment American research university \parencite{aiken2019modeling}. It includes timestamped course grades, demographics, majors declared, degrees awarded, and preparation information for \totalN $\;$students for the years 1992 to 2012. Student's are included in the study if they remain enrolled for at least 5 semesters. The 5 semester mark is chosen because a large number of students(18.3\%) remain enrolled for only four or less semesters taking core courses \parencite{yue2017rethinking}. These students do not graduate the university and likely transfer elsewhere as their institutional commitment may be low \parencite{tinto1975dropout}. 88.02\% students who remain enrolled for at least 5 semesters graduate from this university.

\begin{figure}
    \centering
    \includegraphics[scale=0.65, trim={2cm 0cm 2cm 2cm}]{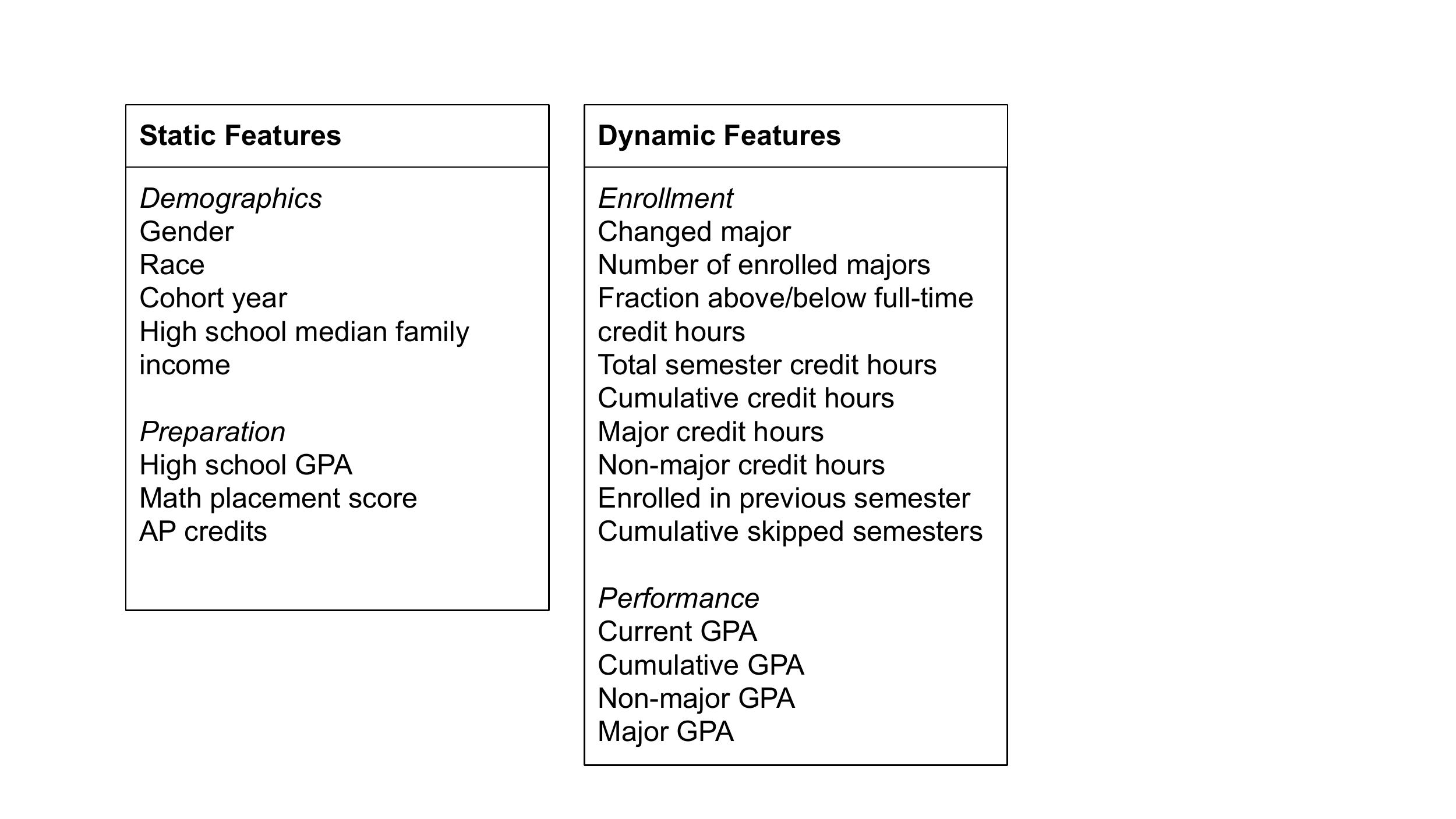}
    \caption{Data model for all models presented in this paper.}
    \label{fig:features}
\end{figure}

In this study, data is organized by semester into of four categories: demographics, preparation, enrollment, and performance (Fig. \ref{fig:features}). Demographics and preparation are considered "static" features. That is, they are not changing over the course of the study. The enrollment and performance features are considered "dynamic". That is, they are cumulative and values can change as students progress in their studies.

The demographics data includes gender, race, cohort year, and the median family income for the zipcode the high school the student attended. The gender category is binary: male or female. The race category uses the IPEDs definitions \parencite{ipedsdefinitions} which is then reduced to a binary category of white/asian or other. This is due to two reasons: 1) the university has a primarily white/asian population, and 2) in this paper we establish model comparison's via out-of-sample prediction accuracy. Thus due to the small minority population, out-of-sample, per semester data may not include all racial groups with sufficient numbers ($>$100, \cite{poldrack2019establishment}). The cohort year is the year the student begins their first courses. The median income comes from the 2011 American Community Survey 5-Year Estimate \parencite{2011census}. If the student has a reported high school GPA, then we also know the zip code of the high school the student attended. This zipcode is matched to the 2011 census data zip codes. The census data includes the median incomes for families living in that zipcode. The log of the median income is then recorded per student. If the zip code data is missing the median income is imputed as described in Section \ref{sec:methods}. This method is not an attempt to provide a course grained estimate of the socio-economic status of the student. Instead, it is a method to obtain a measure of the socio-economic status of the high school the student attended. Student learning can be impacted not only by a student's personal socio-economic status, but also the socio-economic status of the learning community they belong to \parencite{caldas1997effect}. We do not have individual level socio-economic data such as parental income. Nor do we have data on financial aid status.

The preparation data includes the per student reported high school GPA, a math placement score, and if amount of Advanced Placement credits if any. The math placement score comes from a 30 item test that the university requires students to take upon admission. This test has been used for the entirety of the study. A higher score will place the student in a higher math course up to calculus 1. High school GPA is also reported for each student. In both the cases of high school GPA and math placement score the data can be missing for several reasons. High school GPA is not required to be reported by applicants to this university, thus high school GPA is not always recorded for each student. If students have transfer courses for higher math courses (e.g., calculus 1) then the student does not need to take a placement test to determine if they can begin in calculus 1. In the case of missing data the math placement score and the high school GPA are imputed as described in Section \ref{sec:methods}.

The enrollment features are organized per semester. They include whether a student changed their major in that semester, the number of currently enrolled majors, the ratio of credit hours to a full time load (12 credit hours), the total registered semester credit hours, the total cumulative credit hours accrued, the non-major credits the student has registered for in this semester, the major credits hours enrolled per semester, whether the student was enrolled in the previous semester, and the cumulative number of skipped semesters.

The performance features are organized per semester. They include the current semester GPA, the cumulative GPA, the current semester's GPA for courses outside of the student's current major, and the current semester's GPA for courses within the student's current major.

% \begin{table}[]
%     \centering
%     \begin{tabular}{cccc}
%     \hline \\
%       feature name  & variable type & mean & stdev \\
%       \hline \\
%          & & & \\
%          \hline
%     \end{tabular}
%     \caption{Descriptive characteristics of static features.}
%     \label{tab:feature table}
% \end{table}

\begin{figure}
    \centering
    \includegraphics[scale=0.5]{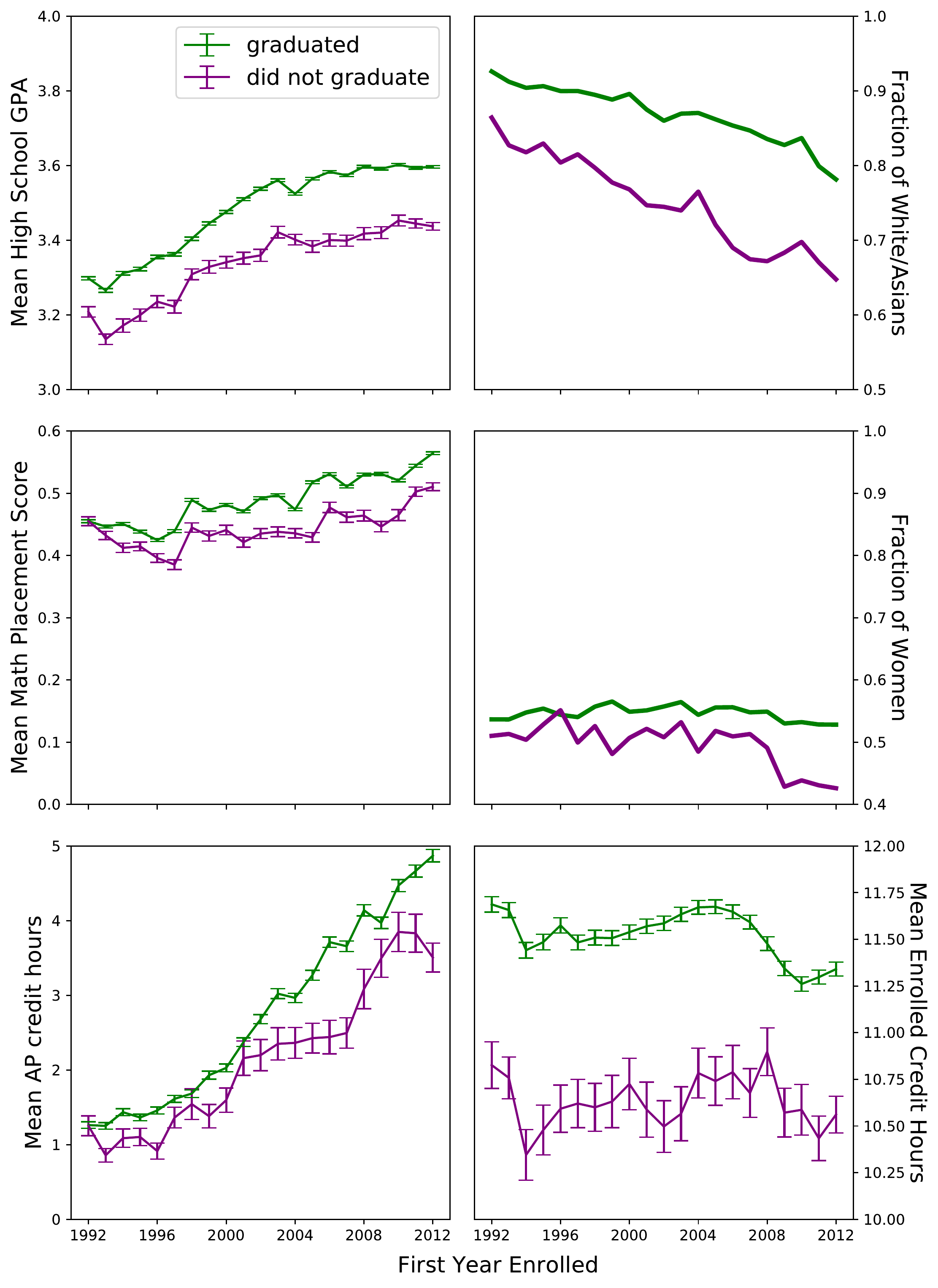}
    \caption{A comparison of a subset of features for students who graduate versus those who do not based on first year enrolled. Since 1992 the student population has become more diverse racially, students have had increasing high school GPAs and math placement scores, and the time it takes students who graduate to graduate has been decreasing. Students who graduate typically take more credit hours than those who don't, typically are better prepared as measured by high school GPA and math placement score, and are less diverse than the total university population.}
    \label{fig:rawdata}
\end{figure}

\begin{figure}
    \centering
    \includegraphics[scale=0.5]{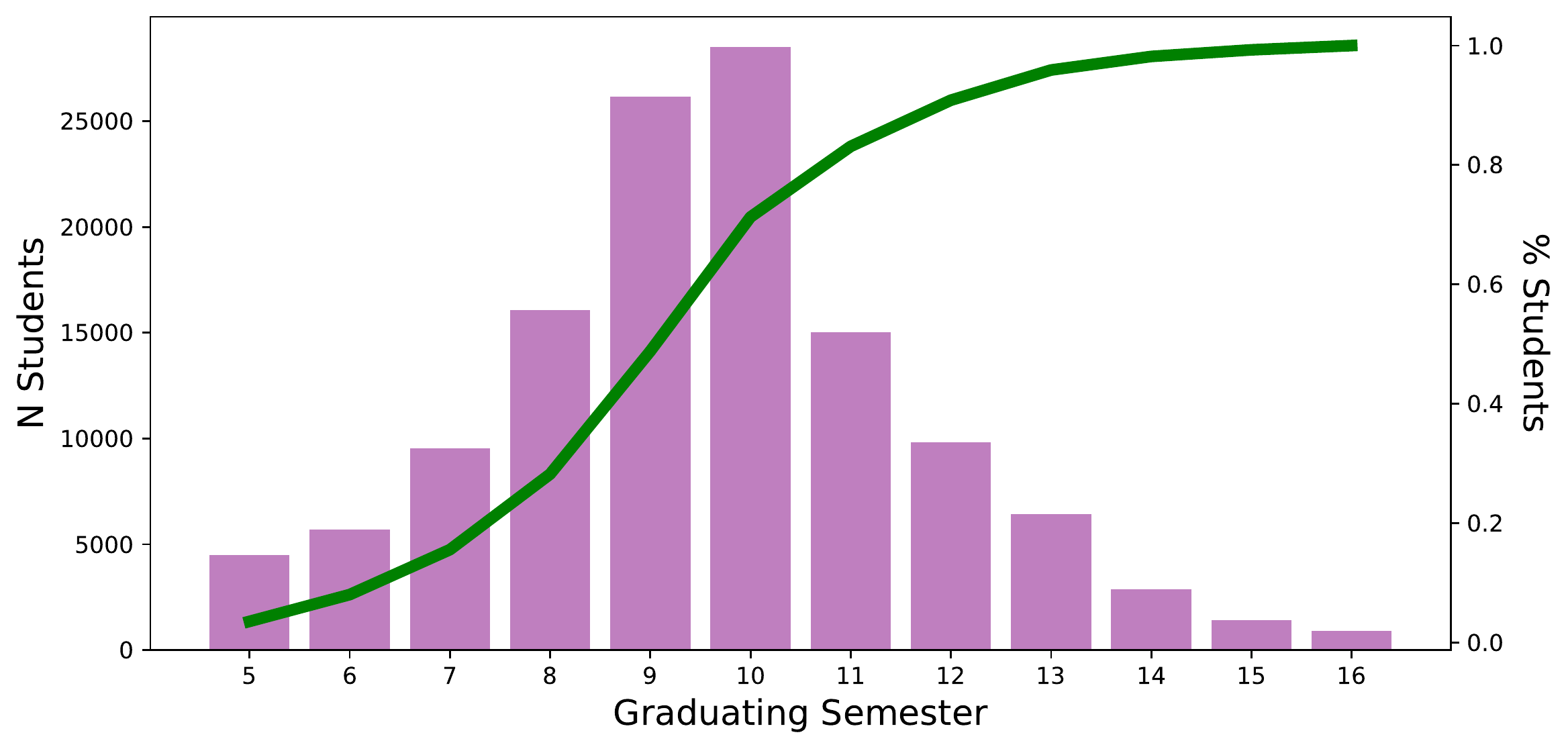}
    \caption{Students typically graduate within the window of 8 to 12 semesters. The bars represent the number of students across all cohort years who graduate in the following semester. The green line represents the cumulative fraction of students who have graduated.}
    \label{fig:gradrates}
\end{figure}
\section{Methods}
\label{sec:methods}
In this paper we have used a discrete time hazard modeling framework for all models that are presented. We begin by comparing the logistic regression model to a gradient boosted (xgboost) \parencite{chen2016xgboost} model to produce the most predictive model. In this section we will describe the discrete time hazard modeling framework, the logistic regression equation, the xgboost model equation, and the statistics and evaluation methods that we have used to determine the xgboost model is the best predictive model. We will then describe the methods we used to evaluate what features the model uses to produce predictions.  The superior xgboost model is then used to describe what factors are most predictive of students graduating in a particular semester.

\subsection{Discrete Time Hazard Models}

Discrete time hazard models are, essentially, logistic regression (or other classification) models calculated per time step for time changing data. They control for time to event predictions when the times to events are simultaneous or explicitly discrete and there is new data being collected over the course of the study period \parencite{yamaguchi1991event}. Traditional hazard analysis such as Cox regression is not capable of doing this type of analysis due to the duration being measured being a common value amongst many students \parencite{yamaguchi1991event}. This is due to the likelihood function of a continuous event duration model assumes independent and unique durations (i.e., the time it takes to graduate). When these durations are not unique, they can lead to overfitting. In this study the time unit used is a semester \parencite{yue2017rethinking}. Thus the models attempt to predict whether a student will graduate in the immediate following semester. If a student drops out from the university, they are not included in following semester data set.

\subsection{Logistic Regression}

The logistic regression equation used in this paper is as follows:

\begin{equation}
    \log \frac{y(t)}{1-y(t)} = \beta_{0}(t) +  \boldsymbol{\beta}_{S}^T \boldsymbol{X}_{S} + \boldsymbol{\beta}_{D}(t)^{T} \boldsymbol{X}_{D}(t)
\end{equation}

Where $y(t)$ is the likelihood to graduate in the following semester and $\boldsymbol{\beta}_{S}$ and $\boldsymbol{X}_{S}$ are static features indicated by subtext $S$ (Fig. \ref{fig:features}). The dynamic terms (indicated by subtext $D$) are calculated per semester $t$ from semester 5 until semester 16 \parencite{yue2017rethinking}. $\epsilon$ is then the irreducible random error. This is effectively an iteratively calculated multinomial logistic regression model \parencite{yue2017rethinking}. The model is fit using the maximum likelihood estimation method. No regularization is used.

The logistic regression model is built using the statsmodels library in python \parencite{seabold2010statsmodels}.

\subsection{Gradient Boosted Trees}

Xgboost is an implementation of a stochastic gradient boosting machine \parencite{friedman2001greedy, friedman2002stochastic, chen2016xgboost} that can also be used to attempt to solve the logistic equation. Gradient boosted machines can be seen as models that are iteratively fit on the current residuals starting from the null model. The additive collection of these models produce the output. Xgboost uses decision trees as its base learners. 

The gradient boosted model is thus:

\begin{equation}
    \log \frac{y(t)}{1-y(t)} = F_{0} + \sum_{m}^{M}h_{m}(\boldsymbol{X}(t))F_{m}(\boldsymbol{X}(t))
\end{equation}

Where $m$ is the iteration index, $h_{m}(X)$ is the previous iterations residual model and $F_{m}(X)$ is the current iteration's model fit to previous iteration's residuals. The total number of iterations is set by $M$.

A gradient boosted machine can use any learner for the iterative procedure. In this paper we use the decision tree learner as implemented in Xgboost \cite{chen2016xgboost}. Decision trees are models that use a tree like structure to fit data and produce regressions and classifications. For each "leaf" node in a tree, a specific decision is made that separates data into two diverging paths. Each node represents a single variable. Categorical variables (e.g., gender) are simply split by the category. Continuous variables (e.g., semester GPA) are split by a decision boundary (e.g., GPA $>$ 3.0). This boundary is typically determined through iteratively fitting the model to find the best boundary.

In Xgboost, each decision tree is trained from a randomly sampled set of rows and columns. Each tree is grown up to a maximum depth using a leaf growing algorithm that estimates whether an additional leaf will produce a better or worse tree \parencite{chen2016xgboost}. Thus there is no separation between static and dynamic features as in the logistic regression model for each semester. All xgboost models are built using the xgboost library in python \parencite{chen2016xgboost}.

Xgboost has a built in algorithm to deal with missing data \parencite{chen2016xgboost}. When missing, data is imputed for each tree by selecting the most common decision path for a node. Because a feature may appear in many trees in the model, the decision boundary can be very different per tree in comparison to an mean imputation scheme.

Xgboost models have hyperparameters that define how the model is to be constructed prior to the model being trained. These hyperparameters govern two classes of model design: 1) how the boosting functions, and 2) how the trees are grown and structured. In the case of boosting, the most common hyperparameter used is the number of total boosting learners. For the tree structure this includes the maximum depth a tree can grow to and the fraction of variable and row sub-selection. Typically these hyperparameters are determined by grid search when the total combination of hyperparameters is below 1000 combinations \parencite{geron2017hands, bergstra2012random}.

% Xgboost models have hyperparameters that define the boundary conditions that effect the model design. Hyperparameters are model parameters that define the meta-properties of a model instead of the input data. In the case of a gradient boosted trees model, these hyperparameters govern two classes of model design: 1) how the boosting functions, and 2) how the trees are grown and structured. In the case of boosting, the most common hyperparameter used is the number of total boosting learners. For the tree structure this includes the maximum depth a tree can grow to, the fraction of variable and row sub-selection, and regularization terms that can be used in training each learner. Typically these hyperparameters are determined by grid search when the total combination of hyperparameters is below 1000 combinations \parencite{geron2017hands, bergstra2012random}.

Because the learners in the xgboost model are decision trees and not linear models, the xgboost model does not report typical coefficient values like in a traditional logistic regression. Instead they report feature "importances" known as "gain". Each time a variable is used in a tree, the tree is built optimally by splitting in the optimum location. The increase in accuracy due to this split is the gain. The feature importance for a specific variable reported then is the average gain across all the instances that the variable is used in in the model. Xgboost uses a custom split finding algorithm that compares the utility of growing a tree to its maximum depth based on increase in accuracy \parencite{chen2016xgboost}.

\subsection{Model Evaluation}

\begin{figure}
    \centering
    \includegraphics[scale=0.75, trim={2cm 0cm 2cm 2cm}]{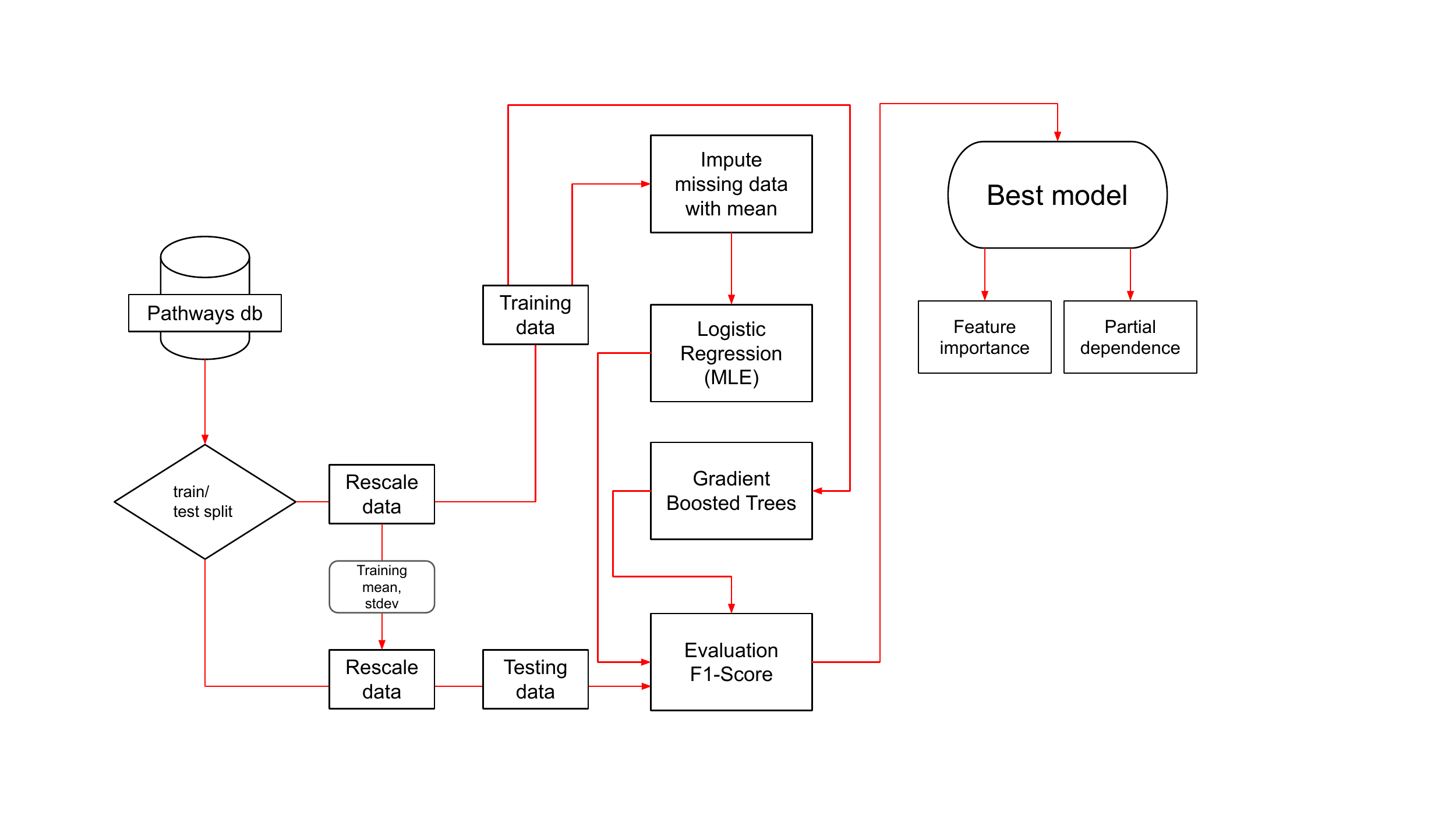}
    \caption{Model evaluation flowchart. Data is split into testing and training sets and is evaluated for two separate models: logistic regression solved by maximum likelihood and gradient boosted trees (xgboost). Missing data for the xgboost model is not imputed from the mean and instead uses the built in imputation engine within xgboost \parencite{chen2016xgboost}.}
    \label{fig:model_eval}
\end{figure}

% \textcolor{red}{paragraph about training testing data and how the integrity of training students vs testing students is maintained across semesters, find a reference on data leakage to motivate this whole thing} \\

The goal of this study is to produce a predictive model of when student's graduate. 
% Two models are employed in this study that attempt to solve the logistic function: logistic regression via maximum likelihood estimation and gradient boosted trees (xgboost, \cite{chen2016xgboost}). 
We use a number of techniques to verify the veracity of the model and increase its accuracy (Fig. \ref{fig:model_eval}). These include: splitting the data into a training/testing sets, imputing missing data, and picking custom thresholds for the predicted probability of graduation. We also limit the over-estimation of a variable's impact on when a student graduates by weighting explanation methods with the $F_{1}$-score \parencite{hastie2005elements}.
% We also have worked to limit the over-estimating of a variable's impact on when a student graduates. To do this, explanation methods such as a model's feature importance or the partial dependence is weighted by the $F_{1}$-score. 
The following section will explain these techniques in detail.

The data processing and model evaluation follows the procedure shown in the flowchart in Fig. \ref{fig:model_eval}. Data for the discrete time hazard model models are:

\begin{enumerate}
    \item queried from the pathways database \parencite{aiken2019modeling}
    \item Continuous values such as high school GPA and semester GPA are rescaled using the z-score. The means and standard deviations are determined by either the starting year cohort (high school GPA) or the timestamp for the enrolled semester (semester GPA). The median income is scaled using the logarithm.
    \item Data is split into training (50\%) and testing (50\%) sets. Since the discrete time hazard models are evaluated per semester data is split by student and not by semester. Thus a student in training data in semester 5 is also in training data for subsequent semesters.
    \item Missing data that is input into the logistic regression model is imputed from the means calculated during the rescaling step.  Missing data for the xgboost model is not imputed prior to fitting the model since the xgboost algorithm has an imputation engine built in.
    \item Each model is trained and evaluated using the $F_{1}$ score. In-sample $F_{1}$ scores are used to determine the best threshold for splitting the predicted probability distributions for classification (i.e., does the student gradute in the next semester).
    \item Models using the selected threshold are then evaluated using out-of-sample test data to compute the $F_{1}$ score.
\end{enumerate}

Several data handling procedures were used to prevent model overfitting, data leakage, and poor predictive performance due to unbalanced class issues. To evaluate the predictive capability of each model, data in this paper is separated into training and test sets for each semester that the discrete time hazard model model is applied. Separating data into partitioned training and testing sets is important to prevent a too optimistic evaluation of the model error\parencite{hastie2005elements}. Without having hold out data to evaluate the model performance, we have no way of knowing how a model will predict the outcome of data it has not seen before \parencite{hastie2005elements, Hofman486, poldrack2019establishment}. Using hold-out data is atypical in recommendations within education research for assessing predictive ability \parencite{raudenbush2002hierarchical, van2019modernizing} and is not used typically in modern papers using discrete time hazard analysis (e.g., \cite{yue2017rethinking}).

Data leakage could occur between two semesters if, for example, a student in semester 6 is in a training set and the same student is in a test set in semester 7 \parencite{poldrack2019establishment}. This is due to some of the features being cumulative thus they would carry forward information from the previous training period into the next test period. Thus, students are identified prior to model training as belonging to the training or testing data set. Students in the test data set are only given to the model to evaluate model predictive performance. 
% Partitioning data into training and testing data sets also helps to prevent model over-fit issues \parencite{hastie2005elements}.

Students do not always have an entry in the database for their high school GPA or their math placement score. This can be due to a variety of reasons such as the high school GPA was not reported, the reported high school GPA was self reported and potentially untrustworthy, and the math placement score was not given to the student because they transferred math credits from another institution. To prevent errors associated with removing data \parencite{rubin1976inference}, this data is imputed in one of two ways. For the logistic model data is imputed from the mean for the student's cohort year. For the xgboost model data is imputed within the model. Xgboost will actively impute missing data within each tree learner based on the most likely branch for each decision node (see Section \ref{sec:methods} and \cite{chen2016xgboost} for more details).

% \textcolor{red}{paragraph about different imputation models: 1) for logistic regression data is imputed by the mean, 2) with xgboost it is imputed within the model itself}\\

% To address this issue we use the synthetic over-sampling minority technique or SMOTE \parencite{chawla2002smote}. SMOTE over-samples the minority class by creating new synthetic examples using the $k$-nearest neighbors to each minority sample. In this paper $k$ is determined by a grid search \parencite{lerman1980fitting} as estimated by the best $F_{1}$ score for the interval [1, 100]. \textcolor{red}{you need to actually grid search the smote k not just say you did it} \\

The predictive ability of models in this paper is estimated using the $F_{1}$-score and the recall \parencite{goutte2005probabilistic}. The $F_{1}$-score is the harmonic mean of the precision and recall and is calculated with the following equation: 

\begin{align}
        F_{1}\text{-score} = 2 \frac{\text{Pr} \cdot \text{Re}}{\text{Pr} + \text{Re}} \\
        Pr = \frac{\text{true positive}}{\text{true positive} + \text{false positive}} \\
        Re = \frac{\text{true positive}}{\text{true positive} + \text{false negative}}
\end{align}

The $F_{1}$ score is over the range 0 to 1 with 1 being perfect performance and 0 being random performance. Precision is defined as the ratio of true positives to the sum of true positives and false positives. That is, it is the ratio of true predicted graduations per semester to the sum of the true predicted graduations and those predicted to graduate but actually do not. Precision is over the range 0 to 1 with 1 being perfect performance and 0 being random performance. Recall is defined as the ratio of true positives to the sum of true positives and false negatives. That is, it is the ration of the number of students predicted to graduate to the total number of people who actually graduated per semester. Recall is over the range 0 to 1 with 1 being perfect performance and 0 being random performance.

We use the $F_{1}$-score over other statistics (such as the area under the reciever operating curve (AUC), \cite{hastie2005elements}) because it balances the trade off between high precision (which includes falsely labeling students as graduated) and high recall (which includes falsely labeling students as not graduating). We use the Recall score in the cases of understanding model performance for sub-groups (such as under-represented minorities) because it gives a more direct interpretation of how accurately the model selects the graduating case.

Students are enrolled for many semesters however they only graduate in a single semester. Thus the majority class in each semester is that a student will \textit{not} graduate in the following semester. This unbalanced class issue can lead to under-fitting of the model specifically such that models simply predict the majority class (in this case not graduating) far too often \parencite{weiss2007cost}. We attempted to use an over-sampling technique \parencite{chawla2002smote} to address this issue however this did not increase model performance. Instead we choose custom thresholds for the model probability distributions for predicting when a student graduated. The custom thresholds are picked using the $F_{1}$ scores calculated from the training data \parencite{guyon2006introduction}.
% Because data is imbalanced (i.e., students do not graduate at equal rates for each sememster), the predicted probability threshold is set by the best in-sample $F_{1}$-score \parencite{guyon2006introduction}. 
The best threshold is calculated via a grid-search between a threshold of 0 to 1 using increments of 0.01. This still, typically, does not increase $F_{1}$ scores.

Beyond predicting when a student will graduate we are interested in the factors that explain why a student was predicted to graduate (or not graduate). A typical logistic regression model using maximum likelihood produces coefficients that represent the magnitude and direction a particular feature given the other variables. For gradient boosted trees these values are calculated differently. Xgboost uses instead the average gain in prediction accuracy across all instances a feature is used in each tree learner. This is commonly called the "feature importance". Additionally we can estimate how much a feature, depending on individual values, contributes to the predicted probability of graduating. In this case we use a method called "partial dependence" that was originally developed to use with gradient boosted models \parencite{friedman2001greedy}. Both the xgboost model and the logistic regression model have a different level of confidence in the prediction per semester (Fig. \ref{fig:f1scores}). Due to this variable confidence our confidence in feature importances and partial dependences varies as well. In order to account for this confidence we have weighted the feature importances and partial dependences with the out-of-sample $F_{1}$ score. This is explained below.

We estimate the effects of model features on prediction using the using the weighted mean of the feature importances. For all 11 semesters the feature importance is weighted by the overall $F_{1}$-score for the semester following this equation:

\begin{equation}
    \boldsymbol{\beta}_{weighted} = \sum_{i=0}^{N=11} F_{1}^{T} \boldsymbol{\beta}_{i}
    \label{equation:weightedfeats}
\end{equation}

This then allows for the cumulative weighted features in Fig. \ref{fig:feat_cumweights} and the weighted features in Fig. \ref{fig:weighted_feats}.

% It is then averaged together with all other feature importances to produce an overall feature importance. Bootstrapped confidence intervals are then calculated for each feature (Fig. \ref{fig:weighted_feats}).

% \textcolor{red}{is partial dependence too much? i dunno maybe. lets see when we get all the other figures and explanations how long the paper is.}

We also estimate the effects of model features on prediction using the partial dependence \parencite{friedman2001greedy}. A partial dependence plot represents the average model output for a single variable ($S$) across the entire feature space ($C$) in the context of all other features \parencite{friedman2001greedy}. This means that for a specific feature $S$, the exact values of $S$ are first, fixed for all rows in the data set for each unique value of $S$. Then the model is calculated per value. The average contribution of a specific unique value to the overall predicted probability of graduation is then considered the partial dependence. Partial dependence is then assumed to be a continuous function over the entire feature space of $S$. Partial dependence is estimated using the following equation as:

\begin{equation}
    \hat{f}_{x_{S}}(x_{s}) = \frac{1}{n} \sum_{i=1}^{n} \hat{f}(x_{S}, x_{C}^{i})
\end{equation}

The partial dependence is calculated for each semester and is weighted by the $F_{1}$-score for that semester (Fig. \ref{fig:cumcredits_pdp}, \ref{fig:cumgrades_pdp}) in a similar fashion to equation \ref{equation:weightedfeats}.

Partial dependence allows the researcher to see the direct contribution to the predicted probability of a given variable value. Because variables often have many values, they can give fluctuating contributions to the predicted probability. In linear models, it is assumed that variables produce linear contributions to the partial dependence \parencite{friedman2001greedy, hastie2005elements} and thus coefficients of a linear model represent unit increases \parencite{hastie2005elements, hosmer2013applied}. For nonlinear models such as gradient boosted trees, this assumption is relaxed and variables can produce nonlinear contributions to the predicted probability.

\section{Results}

In this study we have two broad research questions: 1) exploring the effectiveness of a gradient boosted logistic model in comparison to traditionally solved logistic models, and 2) quantifying the components of Tinto's Theory of Drop Out in the context of who does or does not graduate at this university during the study period. In this section we will first describe the effectiveness of the gradient boosted model in comparison to the traditional maximum likelihood model. Then we will describe the results that are drawn from the gradient boosted model in the context of Tinto's theory.

\subsection{Gradient boosting}

\begin{figure}
    \centering
    \includegraphics[scale=0.5]{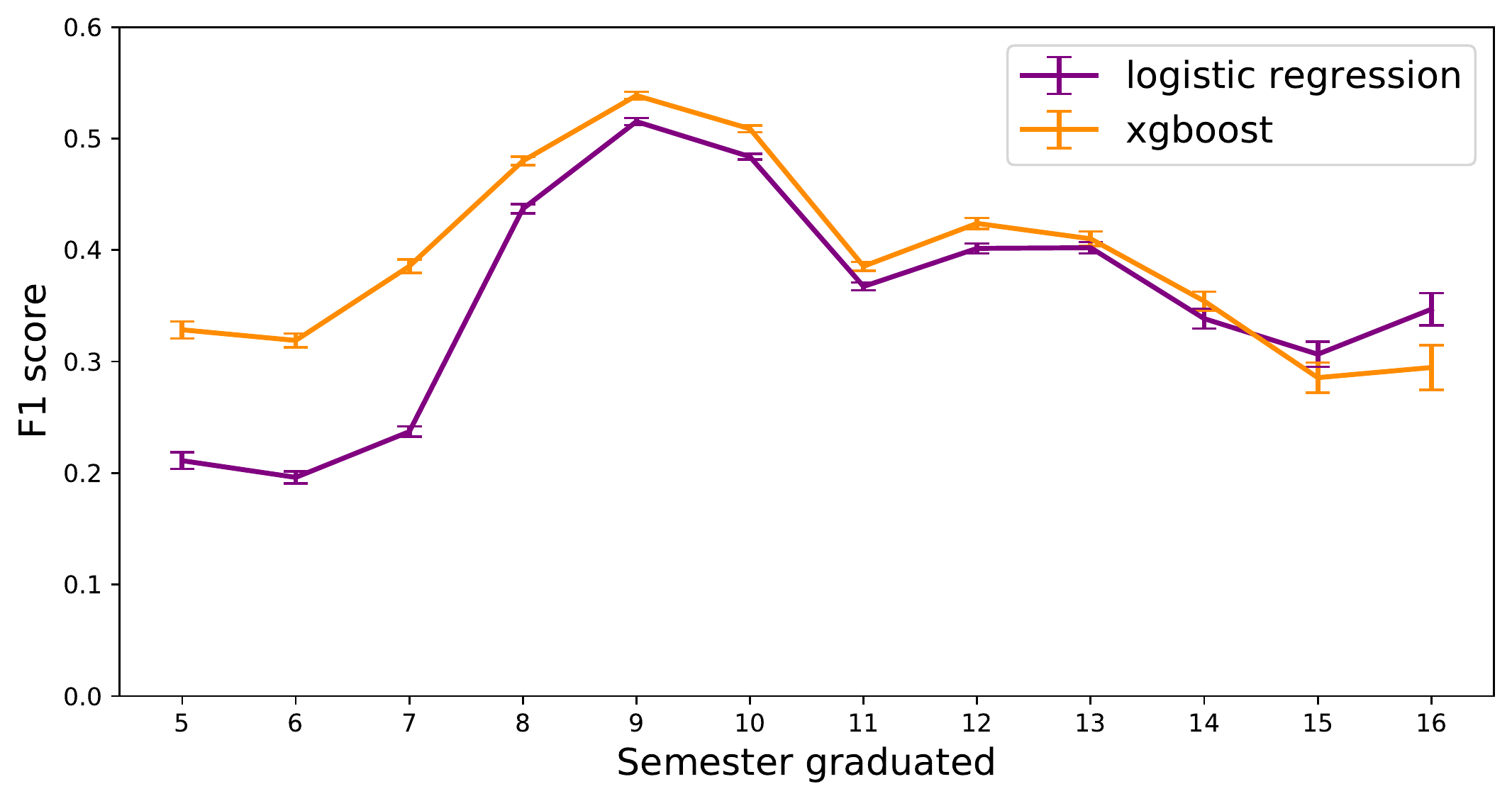}
    \caption{Out-of-sample $F_{1}$ scores for all models per semester. Xgboost clearly increases the precision and recall tradeoff for the bulk of semesters. Error bars are the bootstrapped standard deviation of the $F_{1}$ score.}
    \label{fig:f1scores}
\end{figure}

The models in this study attempt to predict whether a student will graduate or not during the observation period of being enrolled for 5-16 semesters. The xgboost model is generally more effective than the logistic regression model except in the last semesters (15-16) studied as assessed by the out-of-sample $F_{1}$-score (Fig. \ref{fig:f1scores}). Xgboost is particularly more effective in highly imbalanced case of students graduating (approximately 5\% per semester (Fig.  \ref{fig:gradrates})) within $\leq$8 semesters (Fig. \ref{fig:f1scores}). In the more balanced case (semesters 8-14 have approximately 20\% of the students eligible to graduate do so (Fig. \ref{fig:gradrates})), Xgboost still performs better than the logistic regression model (Fig. \ref{fig:f1scores}). In the final semesters (semesters 15-16) logistic regression outperforms xgboost slightly. 

\begin{figure}
    \centering
    \includegraphics[scale=0.3]{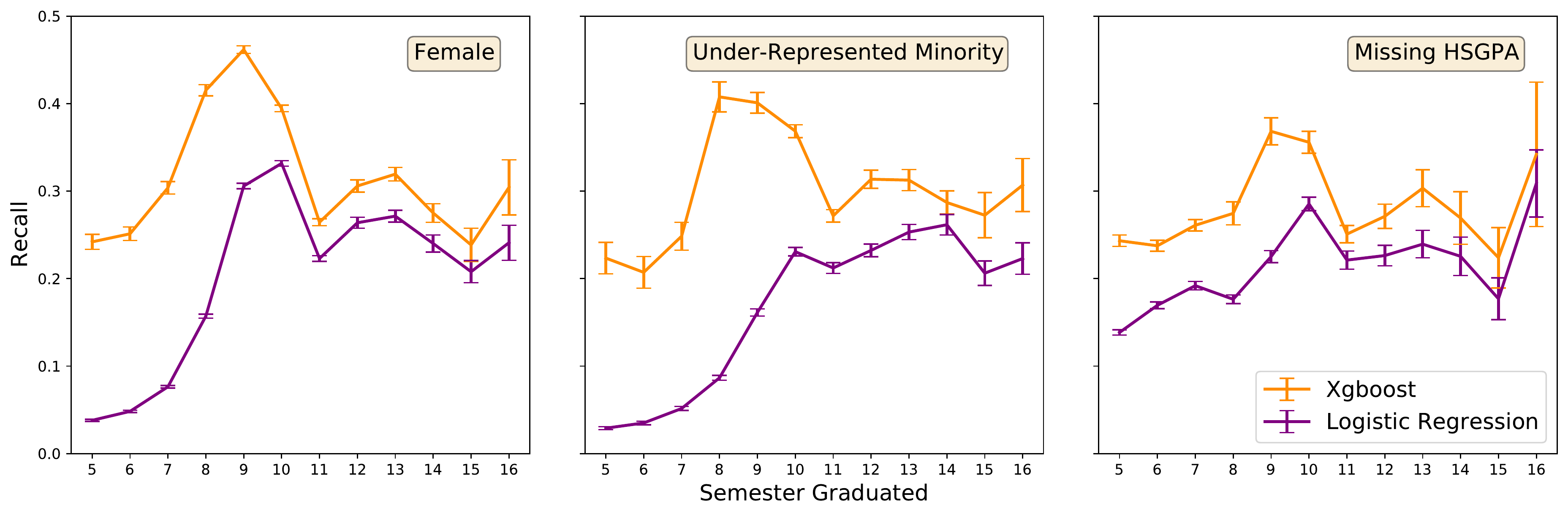}
    \caption{Recall scores for sub-populations within the data. The Xgboost model consistently labels women, under represented minorities, and students with missing data better than logistic regression. In Fig. \ref{fig:f1scores}, there is a large disparity between the logistic regression model and the xgboost model for semesters 5-7. This may be due to the imputation method used with the logistic regression model and the correlation between graduating and missing data during those semesters. However, the gaps in model performance for women, minorities, and missing data in later semesters with no correlation are likely not due to imputation and demonstrate a strong preference for using the xgboost model over the logistic regression model.}
    \label{fig:recalls}
\end{figure}

Xgboost is more effective at correctly predicting the graduating semester of female students as assessed by the out-of-sample recall for all semesters except for semester 16 (Fig. \ref{fig:recalls}). Xgboost is also more effective at correctly predicting the graduating semester of under represented minority  students for all semesters (Fig. \ref{fig:recalls}). Additionally, Xgboost is able to handle missing data cases better than the logistic regression model for all semesters (Fig. \ref{fig:recalls}). Due to the effectiveness of xgboost over logistic regression, the remaining results will focus on the xgboost model.

\subsection{Effects on Time-to-Graduation}

Tinto's Theory of Drop Out posits two overarching quantities as governing student choice to drop out or continue to graduation: 1) educational goal commitment, and 2) institutional commitment. Students initially start university with these commitments. These commitments are then mediated by a student's participating and integration into the academic system and the social system of a given institution. Generally speaking (Fig. \ref{fig:rawdata}), graduating students take more credit hours per semester ($N(grad)=15.1\pm0.01$, $N(no-grad)=13.17\pm0.05$), enroll for more semesters ($N(grad)=10.57\pm0.01$, $N(no-grad)=8.79\pm0.05$), are more likely to be white or asian ($N(grad)=86.7\%$, $N(no-grad)=75.3\%$), and more likely to be female ($N(grad)=54.2\%$, $N(no-grad)=48.9\%$).

\begin{figure}
    \centering
    \includegraphics[scale=0.75]{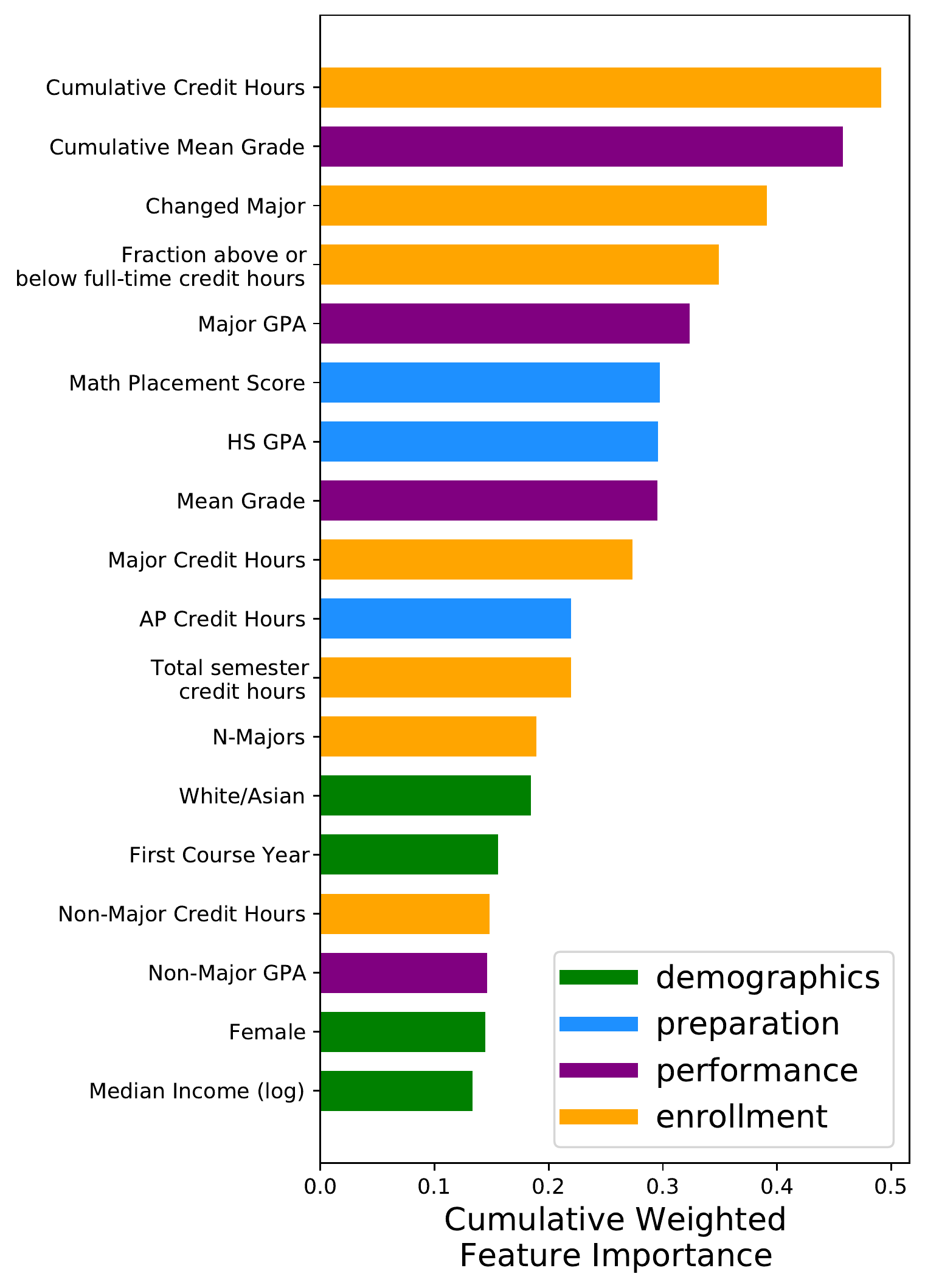}
    \caption{Average feature importance for all graduating semesters for the xgboost model. Feature importances are weighted by the $F_{1}$ scores per semester. Enrollment factors and the cumulative average grade are more likely to predict when a student graduates than other factors.}
    \label{fig:feat_cumweights}
\end{figure}

\begin{figure}
    \centering
    \includegraphics[scale=0.5]{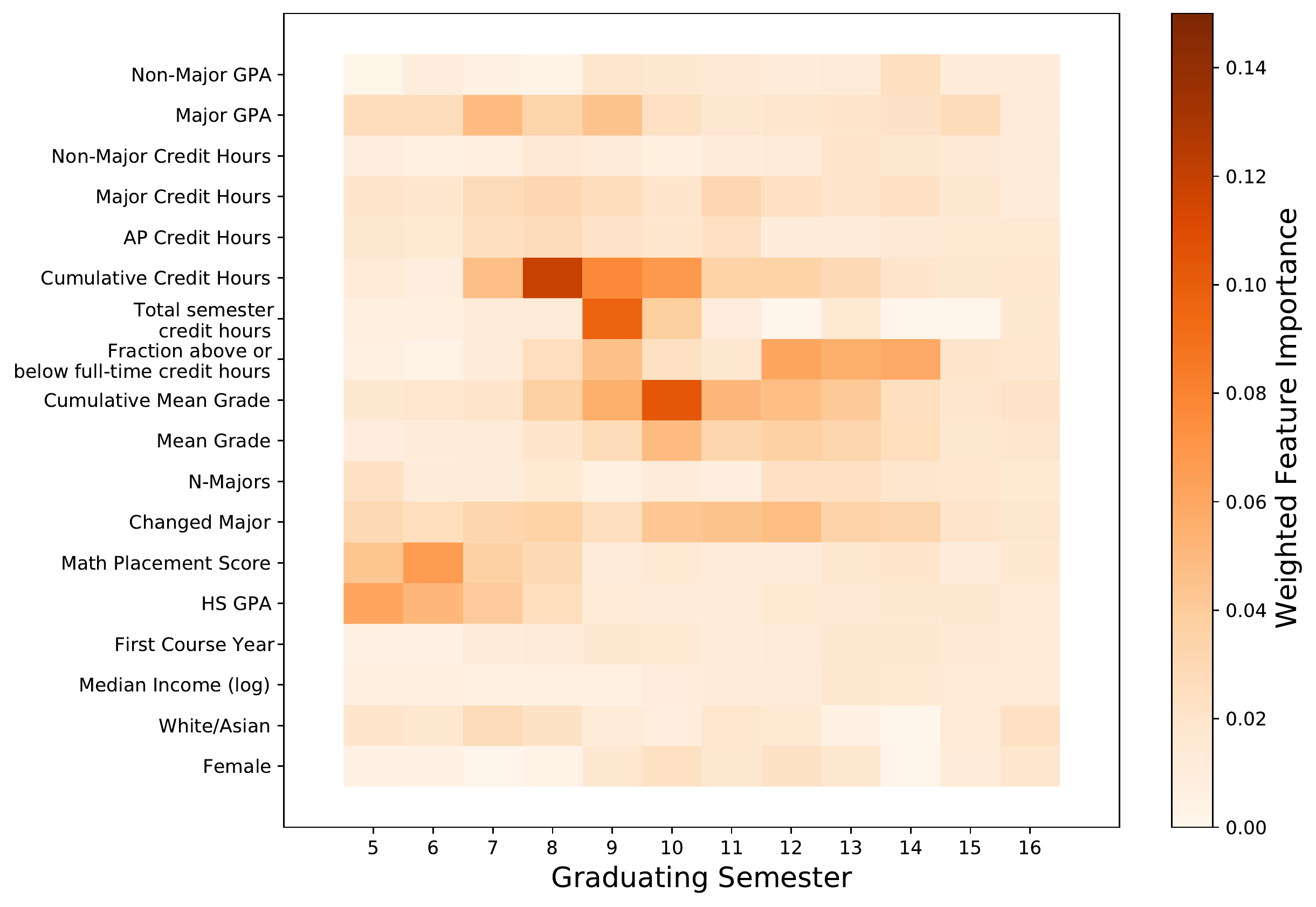}
    \caption{Xgboost feature importances for predicting if a student will graduate in the next semester. Feature importances are weighted by the $F_{1}$ score calculated from test data. The row-wise sum of these features would produce Fig. \ref{fig:feat_cumweights}. By far the most important feature for graduating "on time" (within 8 semesters) is having enough credit hours. Student preparation is important for graduating "early" ($<$8 semesters). Students who change majors are more likely to graduate later and thus this feature becomes more important in later graduating semesters.}
    \label{fig:weighted_feats}
\end{figure}

% initial conditions
In this study, we characterize student's initial commitments as being mediated via the following quantities: high school GPA, math placement score, the number of AP credit hours the student possesses, the median income of the high school the student attended, gender, and race. Preparation is about half as important in predicting when a student graduates in comparison to enrollment factors (Fig. \ref{fig:feat_cumweights}). For early graduating students ($\leq$8 semesters), both math placement scores and high school GPAs were important in predicting that a student will graduate (Fig. \ref{fig:weighted_feats}). While students are likely to have more AP credits on average if they graduate in 8-10 semesters ($AP_{8-10}=4.06\pm 0.03, AP_{other}=2.26\pm 0.01$), having AP credits was not an important indicator for predicting graduation during any semester in comparison to other features (Fig. \ref{fig:feat_cumweights}, \ref{fig:weighted_feats}). While white and asian students and female students are more likely to graduate (Fig., \ref{fig:rawdata}), demographics are not major indicators for predicting when a student graduates for any semester (Fig. \ref{fig:feat_cumweights}, \ref{fig:weighted_feats}). This is also true for the median incomes of the zipcode the students went to high school.

\begin{figure}
    \centering
    \includegraphics[scale=0.5]{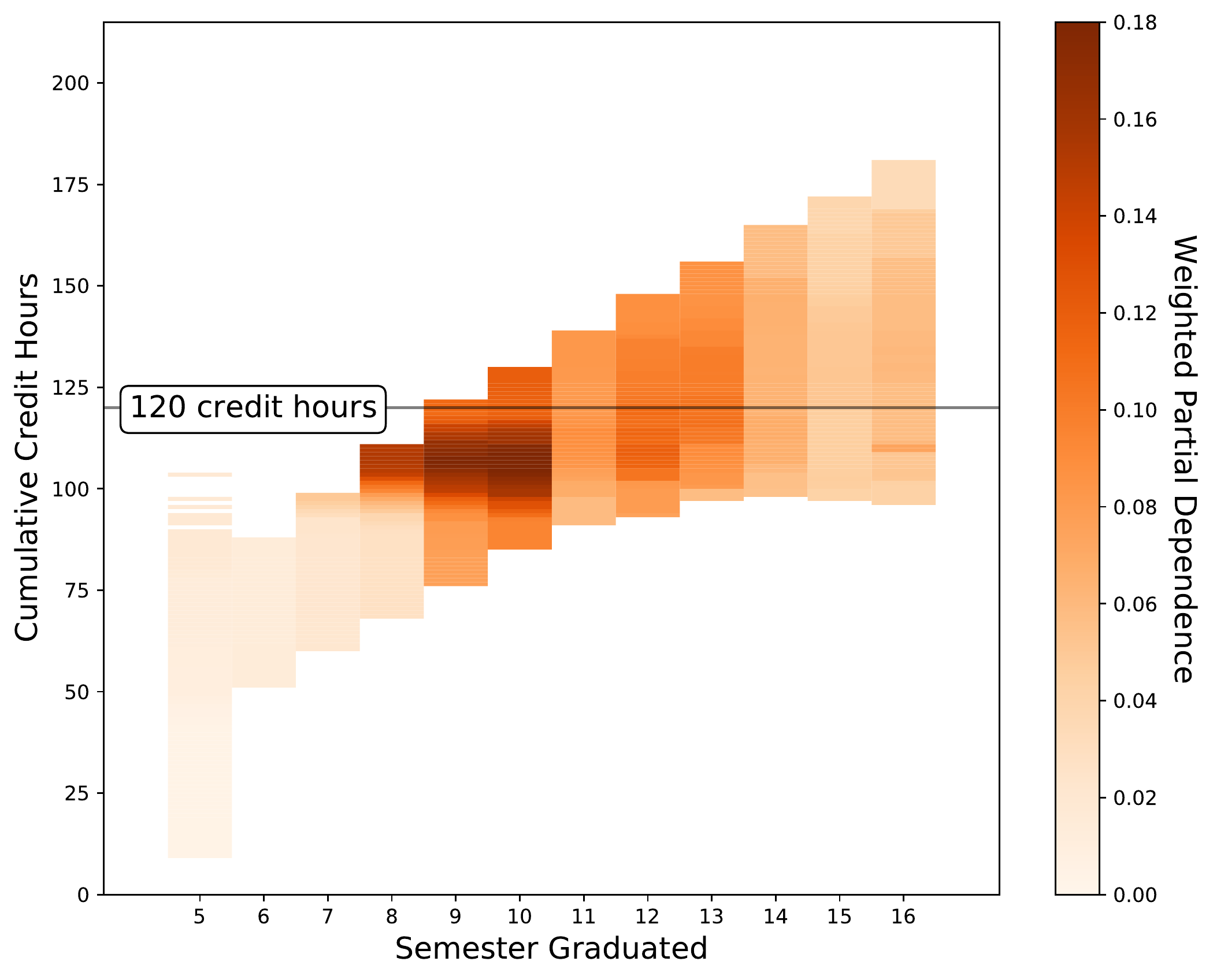}
    \caption{The partial dependence per semester for the cumulative credit hours a student has obtained. The stronger the partial dependence, the more contribution that value of cumulative credit hours has on the predicted probability that a student will graduate in the given semester. The partial dependence has been weighted by the per semester $F_{1}$ score. Having a total number of credit hours close to 120 is highly predictive of graduating if students graduate between 8 and 10 semesters of enrollment. Outside of this window the impact of the total number of credit hours diminishes. This is likely due to students having additional credit hours that do not count towards a degree such as when they change their major.}
    \label{fig:cumcredits_pdp}
\end{figure}

\begin{figure}
    \centering
    \includegraphics[scale=0.5]{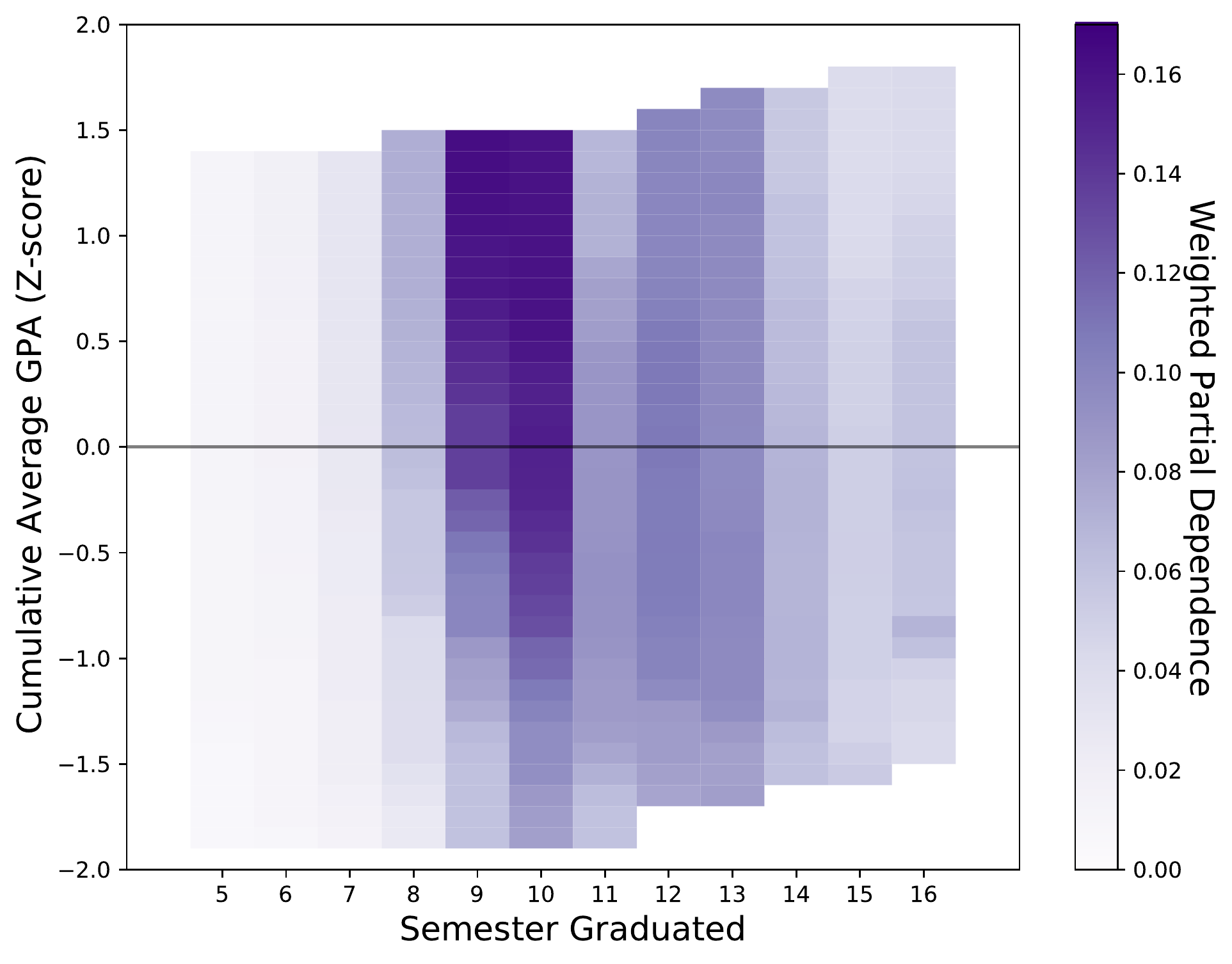}
    \caption{The partial dependence per semester for the cumulative average GPA  a student has obtained. The stronger the partial dependence, the more contribution that the cumulative average GPA has on the predicted probability that a student will graduate in the given semester. The partial dependence has been weighted by the per semester $F_{1}$ score.  Having a far above average GPA is a major contribution to graduating if a student graduates between semesters 9 and 10. This is likely due to these students never failing a course and having a high commitment to their chosen major.}
    \label{fig:cumgrades_pdp}
\end{figure}

% academic integration
In this study academic integration is represented by a combination of features including the cumulative credit hours, cumulative mean grade; and the per-semester fraction above or below full-time credit hour, mean grade, non-major GPA. A student's enrollment and performance are the most predictive features when predicting the semester when a student graduates (Fig. \ref{fig:feat_cumweights}, \ref{fig:weighted_feats}). A student's cumulative credit hours is most important to predicting whether a student graduates "on-time" (after 8 semesters) or not and is overall important for prediction. The cumulative credit hours is less important for prediction the further a student's credits are away from 120 total credit hours (Fig. \ref{fig:cumcredits_pdp}). A student's average grade is most important for predicting that students will graduate within 10 semesters and is overall important for predicting when a student will graduate (Fig. \ref{fig:feat_cumweights}, \ref{fig:weighted_feats}). Having above average grades is important for predicting students who graduate in 9-10 semesters however this is less important for other semesters (Fig. \ref{fig:cumgrades_pdp}).

% social integration

In this study we use per semester major GPA, major credit hours, cumulative number of majors, and whether a student changed major in that semester or not to represent their social integration into the departments and learning communities associated with their chosen degree program. Student's performance within their major was not as an important indicator for predicting graduation in comparison to overall grade and cumulative credit hours (Fig. \ref{fig:feat_cumweights}). Changing a major has increased importance for predicting students who graduate later (Fig. \ref{fig:weighted_feats}).

% \begin{figure}
%     \centering
%     % \includegraphics[scale=0.5]{weighted_feats.pdf}
%     \includegraphics[scale=0.5]{smote_weighted_feats.pdf}
%     \caption{Logistic regression coefficients weighted by the semester $F_{1}$-score, trained on SMOTE data. Error bars represent 95\% confidence intervals. Overall, enrollment features, not performance features, are most predictive of graduation. Changing a major has an overall negative effect on the time it takes to graduate. Preparation as determined by math placement score and high school GPA do not contribute highly to when a student graduates. Being a white or asian student has a positive effect on graduating on time.}
%     \label{fig:weighted_feats}
% \end{figure}

\section{Discussion}

This paper presents two complimentary results: 1) gradient boosting is a useful tool for predicting when students graduate in comparison to traditional statistical algorithms, 2) students who actively integrate into their academic and social communities is the primary effect that predicts when a student graduates thus following Tinto's Theory. This section will discuss why the xgboost model outperforms logistic regression, the implications that come with the results of the model, and compare it to other studies that have predicted time-to-graduation using Tinto's Theory of Drop Out and hazard modeling more broadly.

% why is gradient boosting better than MLE? //

\subsection{Why does gradient boosting produce better fitted models than maximum likelihood estimation?}

In almost every case, the xgboost model predicts graduation better than the logistic regression model. This can be for several reasons. First, the xgboost model fitting procedure is a slower iterative procedure (1000 iterations) than the maximum likelihood estimation (typically 5-6 iterations) that the logistic regression model uses. Thus, in each iteration the xgboost model can focus on the local neighborhood of feature details within the training data that the logistic regression model may miss \parencite{nielsen2016tree}. 

Second, the xgboost model uses a custom imputation engine \parencite{chen2016xgboost}. Whenever data is missing and an learner is using a feature with missing data, the missing data is imputed to be the most likely choice in the decision tree. Because there are many learners, they can account for different local patterns in the data. The data for the logistic regression model is imputed from the mean of the student's starting year cohort. This mean imputation likely loses information that the xgboost model is able to attend to. It could be that a more sophisticated imputation model will increase the logistic regression model performance. Typically if data missingness correlates with the outcome variable, then data should be imputed to prevent bias due to the missing data \parencite{sterne2009multiple}. In our case, data missingness for both high school GPA and math placement score correlates with early graduation rates (see supplemental Fig. \ref{fig:sup_corr}). Thus the logistic regression model performance would likely increase for semester's 5-7 with a more sophisticated imputation model. However this would not explain the substantial improvements xgboost makes over the logistic model for women and under represented minorities in later semesters.

Third, xgboost penalizes leafs within the tree learners that are fit on few examples from the training data \parencite{chen2016xgboost}. Additionally, xgboost weights on class labels as well. This is especially useful in the highly unbalanced case of predicting when a student graduates. Because the weight of training data from students who do not graduate is tuned via a grid search, this prevents the xgboost model from overfitting on the majority class simply due to having more representative samples. 

\subsection{Effects on Time-to-Graduation}

Tinto's theory suggests that students have an initial level of intent to graduate from an institution upon entry \parencite{tinto1975dropout, braxton2000influence}. This intent is the combination of a students family background (e.g., socio-economic status), individual attributes (e.g., academic ability, race), and pre-college experiences (e.g., high school GPA). This intent is then tempered by the student's at-college experience such as social integration \parencite{kezar2014higher}, financial support \parencite{ishitani2003longitudinal}, and academic performance.

\subsubsection{Effects on the initial conditions of educational commitment}

Student's backgrounds can set up a wide variety of contributions to their initial educational and institutional commitments. In this study we assess these initial conditions using a student's high school GPA, math placement score, the number of AP credit hours the student possesses, the median income of the high school the student attended, gender, and race. This university uses a math placement test to determine a student's incoming math ability to place them in the appropriate math course. This test was designed at the university and has been used for the entire study period. This test was not designed with psychometrics in mind. Thus, while the test is representative of some measurement of math ability, it is unclear how much the test is representative of math ability. However, we see that a student's math placement score is the most important attribute that predicts when a student will graduate that is not a performance or enrollment variable \ref{fig:weighted_feats}). This could be similar to \cite{chen2013stem} evidence that the starting math course is very important to staying in STEM. In \cite{chen2013stem}, students taking lower level math courses at college had increased likelihood of leaving the STEM major they were enrolled in for a different major. In our case, students who take remedial math courses simply have more courses to take and thus must remain at the university longer. Given that students in our study who come from higher income communities are able to remain at university longer before graduating, this indicates that remedial mathematics should be examined more in terms of the cost for a student. It should also be noted that in this study we are predicting all students time to graduation even those who are not pursuing quantitative degrees. The result that math ability has a dominant effect in the context of other demographic and preparation variables is consistent with the literature which claims that math ability has an sizeable effect on student performance at university \parencite{gaertner2014preparing}. 

A student's high school GPA is similarly important to a student's math ability in predicting when they graduate (Fig.  \ref{fig:weighted_feats},\ref{fig:feat_cumweights}). This is consistent with some literature that finds that students with higher high school GPAs are more likely to graduate within 4 years \parencite{ishitani2003longitudinal}. \cite{desjardins1999event} found in a similar study that high school GPA had little to no effect on predicting when a student drops out from university. In both this paper and \cite{desjardins1999event} the study uses data from university's that serve primarily students who come from the state the university resides in. Thus this may indicate that the effect of high school GPA on college success is geographically dependent on the quality of high school preparation for university.

Perhaps counter-intuitively having AP credits is not as important as a student's high school GPA or math ability (Fig. \ref{fig:feat_cumweights}). AP credits directly count for college credits. Students with more AP credit have fewer credit hours necessary to graduate. In this study, having AP credits is an indicator that a student will graduate in four years (8-9 semesters). However it is not a strong indicator that a student will graduate early in comparison to a student's math placement score and high school GPA. \cite{desjardins1999event} found that having transfer credits had no effect on on students dropping out. This was also true in a study of physics students who change their majors \parencite{aiken2019modeling}. It could be that AP credits are too random in whether they count for credit or not. A student may choose to take a course anyways they have AP credit for if its in a sequence (e.g., introductory physics) because they may feel ill prepared for the second semester course. It could also be that some students who take AP courses do so with the intent to take a minor or dual major. In this case, the AP courses "free" up more time at university to be able to graduate in four years.

Financial aid is one of the top reasons students leave the university without a degree \parencite{ishitani2003longitudinal}. In this study we do not know if a student has access to financial aid or not. However, we do include the median income for families that live in the zip code of the high school the student attended according to the 2011 census American Community Survey 5- year estimate \parencite{2011census}. This is a rough estimate of the socio-economic status of the high school that the student attends. A course grained measurement like this does not capture all of the nuances of individual students financial support and in our case, we see that this feature was less important in comparison to performance and enrollment features. In our study, we find that coming from higher income high school's is an indicator that a student will take longer to graduate. \cite{caldas1997effect} found that the socio-economic status of a student's peers had an approximately equal effect as a student's individual socio-economic status on high school student exit examination scores. While this effect is small, it may indicate that students from higher income regions have access to more resources and thus can spend more time in college before entering the work force. This is a similar result to \cite{ishitani2003longitudinal} which found that students who come from low income households are likely to graduate within 4 years in comparison to students from higher income households.

\subsubsection{Dynamic contributions to a student's academic integration}

A student's academic integration is defined by Tinto \cite{tinto1975dropout} as being a combination of performance in university and their intellectual development. In this study we characterize performance through both cumulative GPA measurements and per semester GPA measurements. Further, we split this into major and non-major GPAs.

Performance as measured by GPA has had a demonstrated primary effect on graduating \parencite{desjardins1999event, chen2012institutional, yue2017rethinking}. \cite{desjardins1999event} found that there was a decaying impact on drop out due to GPA. The longer a student was enrolled, the less their GPA was likely to be a strongly influencing factor on dropping out. In our study we find somewhat different results, namely students with above average cumulative GPAs are more likely to graduate on time (Fig. \ref{fig:cumgrades_pdp}). However this effect diminishes with time. It could be that this peak effect is due to two reasons: 1) high performing students are more likely to graduate on time \parencite{yue2017rethinking}, and 2) failing a single course significantly sets back both the time to graduate and a student's cumulative GPA. 

There is an interplay between the cumulative credit hours and the cumulative mean grade. In semester 8, the single most predictive feature is cumulative credit hours (Fig. \ref{fig:weighted_feats}). However by semester 10, cumulative credit hours has exchanged the highest rank with the cumulative mean grade. This has a couple implications. First, the decaying impact of grades as noted by \cite{desjardins1999event} in this case begins later at semester 10. Second, it may be that to graduate within 8 semesters (4 years), there are very few paths other than a strictly laid out course schedule with no deviations and no failing grades. Whereas graduating within 5 years allows more leeway for students to take additional courses that could be due to, for example, receiving a minor. 

Second, there is a transition across graduating semesters of what is important (Fig. \ref{fig:weighted_feats}). Early graduation is predicated more by a student's initial conditions than anything else. By semester 8, the overwhelming effect on successfully predicting students graduating in this semester is their cumulative credit hours. Past semester 8, there is a strong combination of features that are predictive of graduating. This suggests that while Tinto's theory indicates that there is a dynamic contribution of on-campus interactions to student educational and institutional commitments, these dynamic contributions may not matter that much for students with very strong backgrounds who wish to graduate early.

% social integration
\subsubsection{Dynamic contributions to a student's social integration}
% something something social integration
A student's social integration is defined by Tinto \cite{tinto1975dropout} as being a combination of peer group interactions and faculty interactions. In this study we use a course grained measurement of peer group and faculty interactions by measuring the total per semester credit hours a student registers for in their major. Additionally, we note when a student changes their major. Changing a major is a distinct situation where a student will leave their peer and faculty group for a new group and thus may indicate low social integration. 

In this study students who attend the university and take a recommended load of courses and pass each course were likely to graduate in 8-10 semesters. When students altered from this path (e.g., a student changes their major) they increased the amount of time it took to graduate or did not graduate during the study period. \cite{chen2013stem} provides, at least for STEM students, a complementary explanation as to why a student may graduate later. This university has a very large enrollment (typically $>$50000 students), a strong research program, and a strong greek life. In each case these may contribute to social integrations given there is more opportunity at this university than some others for meeting new people, participating in research, or participating in social events.

While not being the largest effect, students who take higher amounts of major credit hours were more likely to graduate in semesters 8 and 9 (Fig. \ref{fig:weighted_feats}). This is sensible given's Tinto's theory since taking more major credit hours both works toward's graduation and is an indication of a strong integration into a peer and faculty group. Within the literature, STEM students who took fewer STEM courses in the first year, were enrolled in less challenging math courses in the first year, and performed poorly in their STEM courses in comparison to non-STEM courses were highly likely to switch majors. Tinto would describe these students as having a reduced educational commitment due to not integrating into the social community  and/or lack in the academic engagement of their chosen degree program. In our study we find similar conclusions, higher performance and the number of major credit hours is a more likely indicator of graduating in four years and this effect diminishes over time after the 4 year (8-9 semesters) mark (Fig. \ref{fig:weighted_feats}). Thus, it may be likely that performance in a major and the frequency of major courses taken may be an indicator of lower social and academic engagement described by Tinto \cite{tinto1975dropout}.

In this study we have highlighted that changing a major can have a profound effect on the time it takes to graduate. Changing a major impacts a student's time-to-graduation and is predictive of students who graduate later (Fig. \ref{fig:weighted_feats}). A student who changes their major could do so due to low academic integration \parencite{chen2013stem}. However this low academic integration could represent low social integration as well \parencite{seymour2000talking}. Student's who perform poorly may ask themselves if they "belong" in a major. In many cases STEM students who demonstrate high academic performance change their major due to reasons associated with social and personal interactions at university \parencite{seymour2000talking, chen2013stem}. This especially affects women and under represented minorities \parencite{seymour2000talking}. In this study changing a major does not show a strong correlation with a student's race or gender ($\rho_{race}=-0.07, \rho_{gender}=-0.03$). In many cases, the experiences that lead to changing a major are more nuanced than a single variable can contain thus while race and gender are a strong component to major change as reported in qualitative interviews \parencite{seymour2000talking}, this may not be captured in binary variables. Ultimately, changing majors has been a poorly investigated and deserves more investigation. Future work will look at using Tinto's theory as a framework for investigating student's changing their major.

\subsubsection{Limitations on this Study}

Given that a perfect $F_{1}$-score is 1 and in this paper we report $F_{1}$-scores around 0.5 at the maximum, it is likely that some of these effects such as financial aid or social integration have large effects on when a student graduates.  It is common in some studies, (e.g., \cite{yue2017rethinking}) to use individual heterogeneity models or "frailty" models to assess unmeasured contributions to graduation \parencite{vaupel1979impact}. In our case, we attempted initially to use gamma distributed frailty terms \parencite{yamaguchi1991event, hougaard1995frailty}. These proved to provide no increase in model performance thus were removed from subsequent analysis. Future work will consider what contributes to the unobserved heterogeneity (e.g., direct measurements of peer social engagement such as social network centrality).

Additionally, this paper has focused on a student's educational commitment (per \cite{tinto1975dropout}) as opposed to their institutional commitment. Much of this is mediated by the fact that 88\% of students in the study do in fact graduate, there may be a subset of students who leave simply due to the fact they become disillusioned with the institution and want to pursue a degree elsewhere. Thus there is likely other factors that are not observed that effects when a student a graduates. Many of these factors, such as life events such as family hardship, are necessarily hard to observe for an entire university population. There are also likely factors that are particular to this university that are not common or not impacting at other institutions. Future work will investigate the effect of student social network belonging and the amount of financial aid available to the students.

\section{Conclusion}

This paper has presented a discrete time hazard model predicting when a student will graduate. It uses a novel method to calculate the logistic regression called gradient boosting which has been shown to provide better fits than traditional maximum likelihood. While using more sophisticated imputation with the logistic regression model might have increased performance, the ease of use of xgboosts built-in imputation engine provides a strong advantage in it's use to researchers. Given the utility of gradient boosting in this setting, especially in providing better predictions for under-served populations, this paper recommends that this method be more prevalent in the education research community. Additionally, other methods should be examined such as artificial neural networks \parencite{gonzalez2002artificial}. 

Additionally, this paper used the partial dependence method to examine two of the model variables. Partial dependence tells us the contributions to the predicted probability of the model for the entire feature space. This method allows us to examine the entirety of continuous variables such as GPA instead of reducing them to a single value associated with a model coefficient. This method too should see much broader use in the education research community.

This paper follows Tinto's theory of drop out that predicts social integration and college participation are more likely to impact a student's commitment to graduation than second order effects such as preparation. Future work will connect student academic performance, participation in course work, with social network metrics and financial aid information. Future work will also examine the specific effects that remedial math have on student retention in a major, their time to graduation, and if participating in remedial math courses lowers the likelihood to graduate.

\section{Acknowledgements}
This project was supported by the Michigan State University College of Natural Sciences including the STEM Gateway Fellowship and the Lappan-Phillips Foundation, the Association of American Universities, and the Norwegian Agency for Quality Assurance in Education (NOKUT), which supports the Center for Computing in Science Education. This project has also received support from the INTPART project of the Research Council of Norway (Grant No. 288125). Morten Hjorth-Jensen is supported by the U.S. National Science Foundation (Grant No. PHY-1404159).

\newpage
\section{Supplemental}

\begin{figure}[!ht]
    \centering
    \includegraphics[scale=0.5]{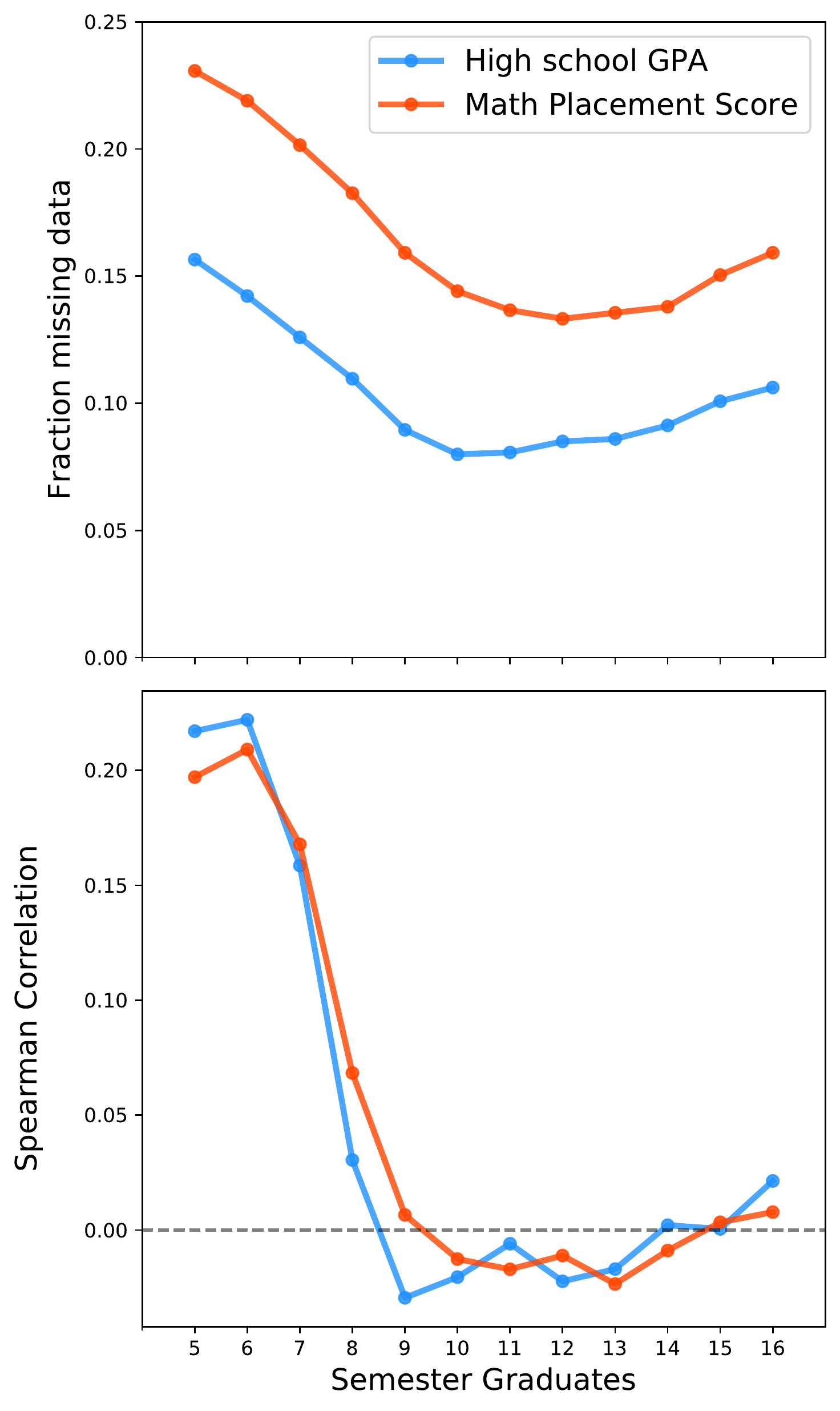}
    \caption{The fraction of missing data and the Spearman rank correlation between whether data is missing and the outcome variable of graduating the following semester. There is a small correlation in the early semesters (5-7) between whether students are graduating and if they have missing data or not.}
    \label{fig:sup_corr}
\end{figure}

\printbibliography

\end{document}